\documentclass[11pt, a4paper]{article}
\usepackage{moriond,epsfig}

\def\be{\begin{equation}}
\def\ee{\end{equation}}
\def\bea{\begin{eqnarray}}
\def\eea{\end{eqnarray}}

\begin{document}
\vspace*{4cm}
\title{ELECTROWEAK SYMMETRY BREAKING WITHOUT A HIGGS BOSON\\
 AT THE LHC}
\author{ Sarah Allwood-Spiers}
\address{{\rm On behalf of the ATLAS and CMS Collaborations} \\
University of Glasgow, Glasgow G12 8QQ, UK.}
\maketitle\abstracts{ We present two studies into strong symmetry
  breaking scenarios at the LHC. The first case is a study into vector
  boson scattering at ATLAS. This uses the framework of the
  Electroweak Chiral Lagrangian with Pad\'e unitarisation to generate
  possible signal scenarios. Signals could be observed with
  an integrated luminosity of
  $\int$Ldt $\simeq$ 30~fb$^{-1}$. Secondly a search for the technirho, $\rho_{TC}$, at
  CMS is presented, within the Technicolour ``Straw Man''
  model. 5$\sigma$ discovery is possible starting from
  $\int$Ldt$\simeq 4$~fb$^{-1}$.}
\noindent
\section{Introduction}
It is possible that the higgs boson does not exist, and that a
 weakly-coupled model is not responsible for electroweak symmetry
 breaking. 
An alternative is that electroweak symmetry breaking results from new strong interactions.
Since the Goldstone bosons resulting from spontaneous symmetry
breaking become the longitudinal components of the $W$ and $Z$ bosons at high energy,
we can probe the electroweak symmetry breaking sector by studying
vector boson interactions.

Strong electroweak symmetry breaking scenarios can be treated quite generally  by
an effective Lagrangian approach, using the Electroweak Chiral
Lagrangian accompanied by some unitarity constraints. A study of
vector boson scattering using
this framework at ATLAS is presented in
section~\ref{sec:EWCHL}. Under the more specific Technicolour ``Straw
Man'' model, a search for the technirho, $\rho_{TC}$, at CMS is presented in section~\ref{sec:TC}.

\section{Electroweak Chiral Lagrangian Studies at ATLAS}\label{sec:EWCHL}
The Electroweak Chiral Lagrangian~\cite{ewchl} (EWChL) describes electroweak
interactions at energies less than 1~TeV. It is built as an expansion in the
Goldstone boson momenta.
If it is assumed that custodial symmetry is conserved, there are only
two, dimension-4, terms that describe the quartic couplings of the
longitudinal vector bosons
\begin{equation}\label{eqn:EWCHL}
{\cal L}^{(4)} = a_{4}(Tr(D_{\mu}UD^{\nu}U^{\dag}))^{2} +
a_{5}(Tr(D_{\mu}UD^{\mu}U^{\dag}))^{2}
\end{equation}
where the Goldstone bosons $\omega_{a}$ ($a$=1,2,3) appear in the group element
$U = e^{\left( i \frac{\underline{\omega}.\underline{\sigma} }{v} \right) }$,
$\sigma$ are the Pauli matrices and $v=246$~GeV. 
Hence the low-energy effect of the underlying physics in
vector boson scattering is parameterised by the coefficients $a_{4}$ and $a_{5}$.

The Lagrangian does not respect unitarity. To extend its validity range to the higher
energies that we will be probing at the LHC, a
unitarisation procedure must be imposed, which can lead to resonances
developing in [$a_{4}$, $a_{5}$] space. This is dependent on the
chosen unitarisation procedure; in the work presented here the
Pad\'e or Inverse Amplitude method was used~\cite{pade}.

There have been several studies of EWChL signals in vector
boson scattering at
ATLAS. 
All seek to exploit the distinctive
characteristics of the vector boson fusion process. The
boson-boson centre-of-mass energy of interest is $\sim$1~TeV, so the bosons
have high-$p_{T}$.
 There are two
high energy forward tag jets originating from the quarks that emitted the
bosons. Since vector bosons are colourless, there is no colour
connection between the tag quarks and hence no additional QCD
radiation in the central region.
\subsection{WW Scattering: $qqWW \rightarrow q'q'WW$}\label{sec:WW} 
An analysis of $WW \rightarrow l \nu qq$ using the ATLAS fast simulation, ATLFAST, to
simulate the effects of the detector is presented
here~\cite{mythesis,stathisthesis}. Five signal points in [$a_{4}$,
  $a_{5}$] space are chosen; after unitarisation these result in a
scalar resonance with a mass of 1~TeV (A), a vector resonance of
1.4~TeV (B), a vector of 1.8~TeV (C), a double resonance of a scalar and a
vector (D), and a continuum scenario (E). This final no-resonance scenario is the most pessimistic, with a cross-section$\times$branching ratio of 13~fb.
Pythia~\cite{pythia}, modified to
include the EWChL, is used to
simulate the signal and the $W+$jets (where $W\rightarrow l \nu$) and $t\bar{t}$ backgrounds. 

The
leptonically-decaying $W$ is reconstructed from the highest-$p_{T}$
lepton and the missing transverse energy, $E_{T}^{miss}$.
The lepton 4-momentum, $E_{T}^{miss}$ and $W$ mass constraint yield a quadratic equation
for the $z$-component of neutrino momentum, $p_{Z}^\nu$. The minimum $p_{Z}^\nu$ solution is chosen because it is closest to the true $p_{Z}^\nu$ in the majority of cases. A cut of
$p_{T}>320$~GeV is made on this $W$ candidate.
\begin{figure}[t]
\centering
\includegraphics[width=15cm]{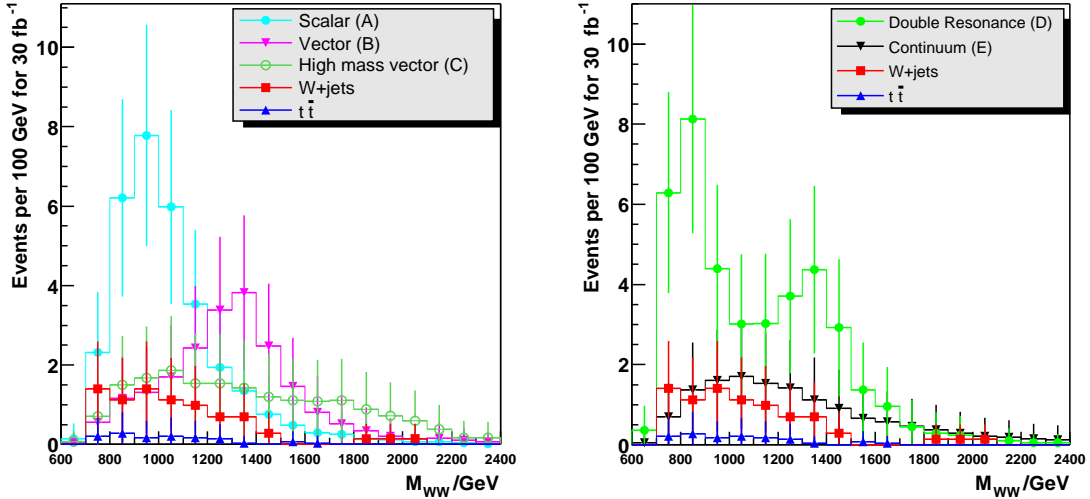}
\caption{Reconstructed $WW$ mass for 5 signal scenarios after all
  cuts.
\label{fig:mww}}
\end{figure}

 The hadronically-decaying $W$ is
highly boosted and can be identified
as one or two jets. 
When jets are identified using the $k_{T}$ algorithm~\cite{ktalg}, the
highest-$p_{T}$ jet is chosen as the hadronic $W$
candidate. It is required to have $p_{T}>320$~GeV and a mass
close to $m_{W}$. A further ``subjet''
cut is performed. The $k_{T}$ algorithm is re-run in subjet mode
over the constituents of this jet and the scale at which the jet is resolved into two subjets, $y_{21}p_{T}^2$, is
found~\cite{WWscattering}. For a
true $W$, this scale is close to $m_{W}^{2}$. A cut requiring $1.55 < {\rm log}(p_{T}\sqrt{y_{21}})<2.0$ reduces the $W+$jets background.

To reduce the $t \bar{t}$ background, a crude reconstruction of tops
is performed by combining
either $W$ candidate with any other jet in the event. Events
in which the invariant mass of any of these combinations is 
close to $m_{t}$ are rejected.
The two tag jets are identified as the highest-$p_{T}$ jets forward and
backward of the $W$ candidates, and required to have $E>300$~GeV and
$|\eta|>2$. 
The $p_{T}$ of the 
full system should be zero, so events with
$p_{T}(WW+
{\rm tag jets})>50$~GeV are rejected. 
Finally, events containing more than one additional central jet with
$p_{T}>20$~GeV are rejected.

The reconstructed $WW$ mass after all cuts is shown in
figure~\ref{fig:mww} for the five chosen signal scenarios. All signals are observable above the
$W+$jets and $t\bar{t}$ backgrounds with an integrated luminosity of
  $\int$Ldt $\simeq$ 30~fb$^{-1}$, with the continuum signal achieving a significance of
$s/\sqrt{b}=4.7$. 
\subsection{WZ Scattering: $qqWZ\rightarrow q'q'WZ$}
\begin{figure}[ht]
\centering
\includegraphics[width=6.7cm]{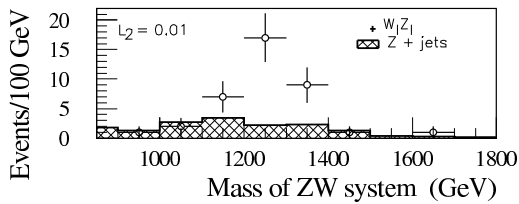}
\caption{Reconstructed $WZ$ mass for $WZ \rightarrow jjll$ after
 all cuts for $300~{\rm fb}^{-1}$.\label{fig:mwz}}
\end{figure}

A 1.2~TeV vector
resonance in $WZ$ scattering with $WZ\rightarrow jjll$ (which has $\sigma \times BR = 2.8$~fb) was investigated using ATLFAST. The analysis considerations are
similar to the above $WW$ study. 
although a different
implementation of cuts is chosen. 
After all
 analysis cuts the only significant background is from $Z+$jets
 production: for 100~fb$^{-1}$, 14 signal events and 3 background
 events are expected in the peak region~\cite{atlasmiagkov}. The reconstructed $WZ$ mass
 is shown in figure~\ref{fig:mwz}.
A recent study using the ATLAS full detector simulation verifies this
result, and also finds that significant signals can be observed with
$100$~fb$^{-1}$ in the $WZ \rightarrow l \nu qq$ mode and
300~fb$^{-1}$ in the $WZ \rightarrow l \nu ll$ mode~\cite{atlasazuelos}.
Updated $WW$ and $WZ$ scattering analyses will be presented in the
forthcoming ATLAS ``Computing System Commissioning'' note to be
completed in summer 2007.

\section{Search for the technirho, $\rho_{TC}$, at CMS}\label{sec:TC}
The original model of Technicolour (TC) is a scaled-up version of
QCD; a new set of interactions is introduced with the
same physics as QCD, but at an energy scale $\Lambda_{TC} \sim 200$~GeV. The
 new strong interaction emerging at the
electroweak scale is mediated by $N_{TC}^2-1$
technigluons. Electroweak symmetry breaking results from the formation of a
technifermion condensate, producing Goldstone bosons (the technipions).
Three of the technipions become the longitudinal components of the
$W^{\pm}$ and $Z$ bosons.

To generate fermion masses, ``Extended Technicolour'' interactions are
introduced, and the technicolour gauge coupling is required to vary
more slowly as a function of the renormalisation scale (it is a
``walking'' rather than a running coupling). The result is that many
technifermions are predicted, and the lightest technicolour resonances
appear below 1~TeV. Acquiring the correct top quark mass is a further
complication; this is achieved by Topcolour-Assisted Technicolour.

The Technicolour ``Straw Man'' model sets the framework for searching
for the lightest bound states. assuming that these can be
considered in isolation~\cite{strawman}. Here we present a search for
the colour-singlet
$\rho_{TC}$ in this framework using the CMS detector.
The analysis~\cite{technistudy} considers the channel $q\bar{q}\rightarrow \rho_{TC}
\rightarrow WZ$ for 14 signal points in [$m(\rho_{TC})$,$m(\pi_{TC})$] space.
 The cleanest decay mode, $\rho_{TC}\rightarrow WZ \rightarrow l \nu
 ll $ is chosen. The $\sigma \times \rm{BR}$ for these signals range from 1~fb to 370~fb.

The main backgrounds are from $WZ\rightarrow l \nu ll$ and $ZZ\rightarrow
 llll$, $Zb\bar{b}\rightarrow ll+X$ and $t\bar{t}$. All signals and
 backgrounds are generated using Pythia~\cite{pythia}. The CMS fast simulation
 FAMOS is used, with lepton reconstruction efficiencies and
 resolutions validated against the GEANT-based full detector simulation.

The three highest-$p_{T}$ leptons (electrons or muons) in the event
are selected. Making
appropriate isolation cuts in the initial identification of these lepton candidates is important in reducing the $Zb\bar{b}$
and $t\bar{t}$ backgrounds. 
The $Z$ is reconstructed from two same flavour opposite
sign leptons. The $W$ is reconstructed from the third lepton and
$E_{T}^{miss}$, as explained in section~\ref{sec:WW}. 

\begin{figure}
\vfill
\centering
$\begin{array}{cc}
\includegraphics[width=7.5cm]{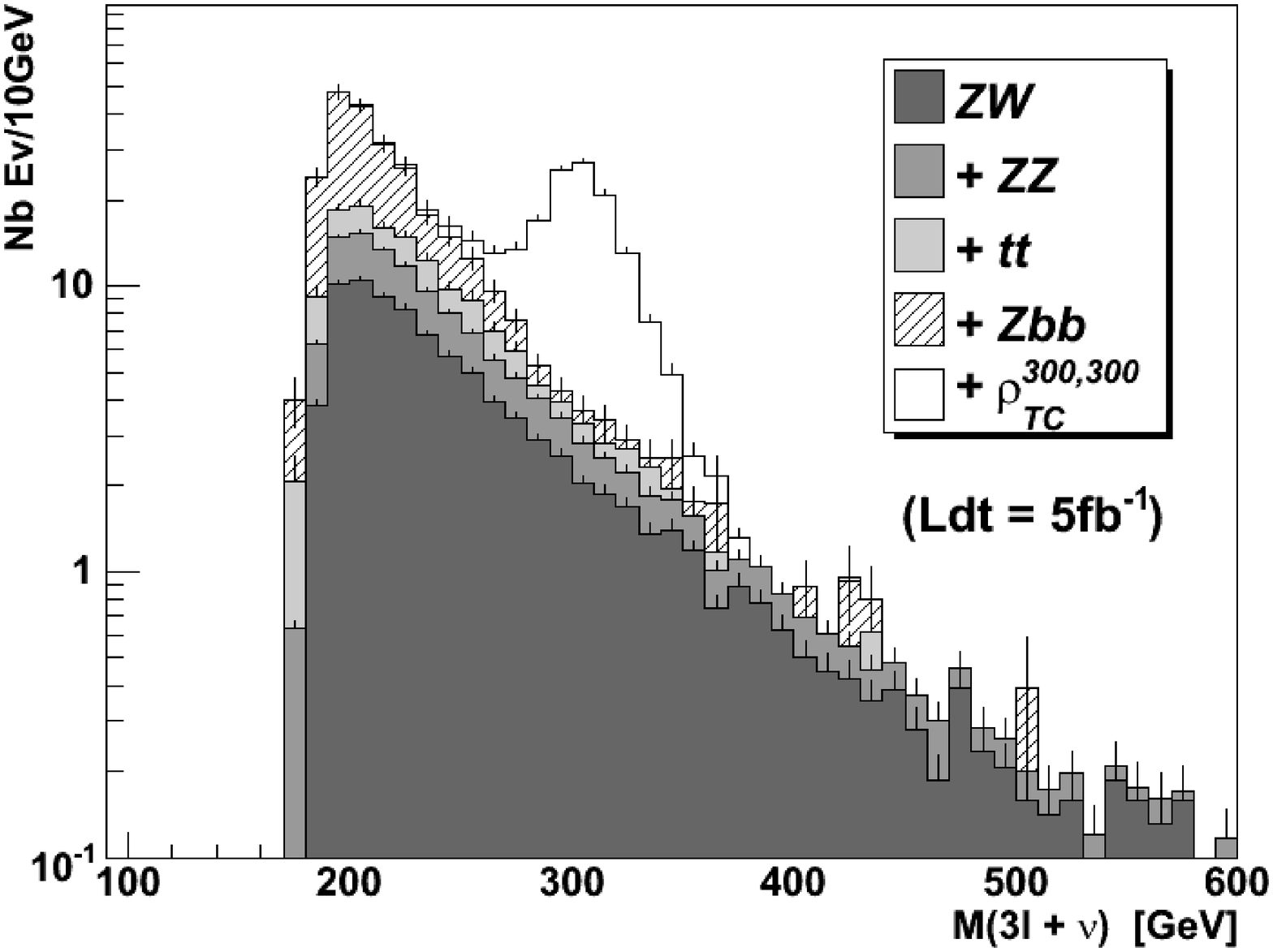}
\hspace{0.1cm} &
\includegraphics[width=7.5cm]{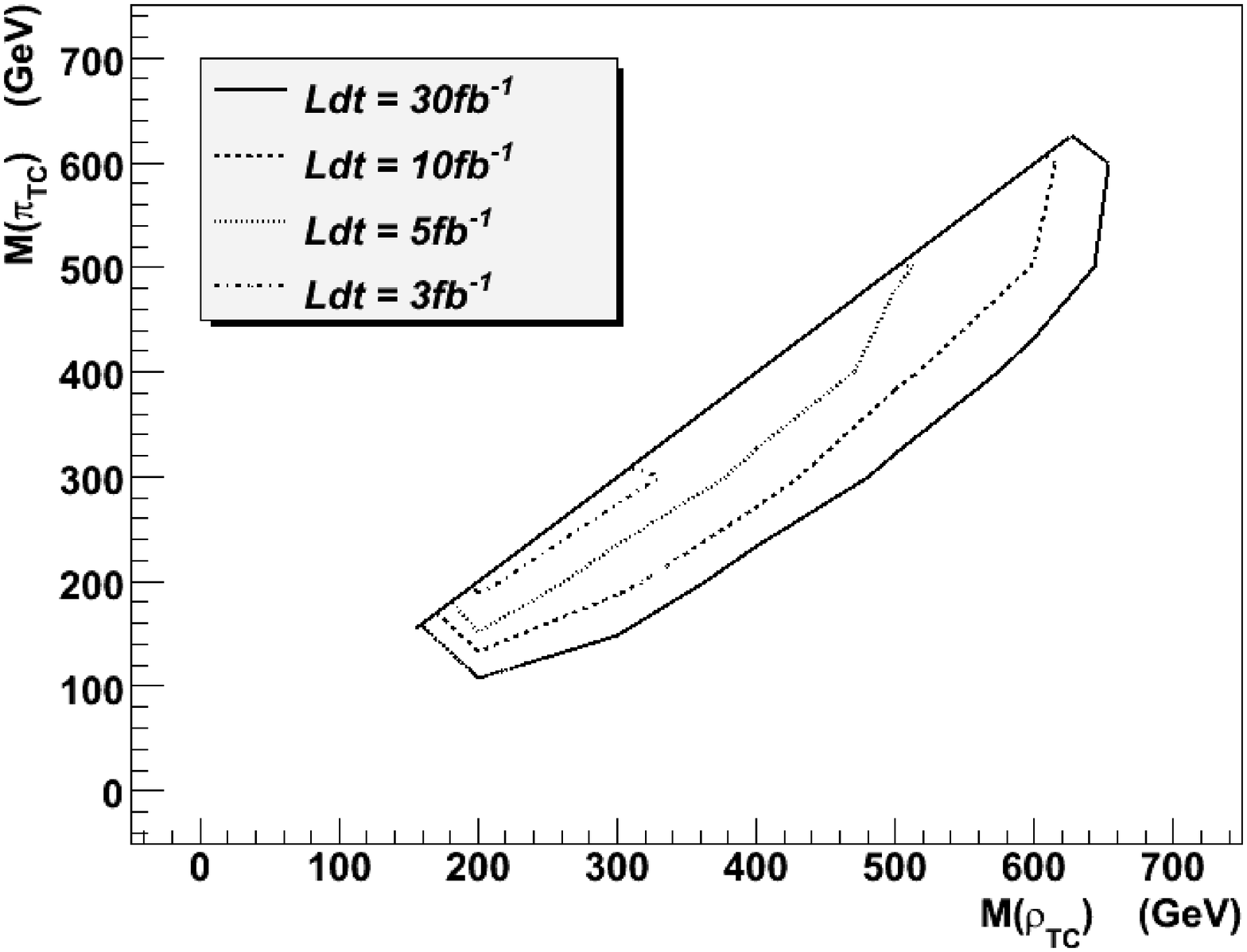} \\
\end{array}$
\caption{(left) Reconstructed $\rho_{TC}$ mass after all cuts, (right) Sensitivity contours for 5$\sigma$ discovery of $\rho_{TC}$ at various integrated luminosities, assuming the default parameters of the TC Straw Man model. 
\label{fig:TCsens}}
\vfill
\end{figure}

Kinematic cuts on the $W$ and $Z$ candidates are needed to improve the signal to background ratio. The $W$ and $Z$ candidates are each required
to have $p_{T}>30$~GeV. A
$Z$ mass window cut of $|m_{l^+l^-}-m_{Z}|<3 \sigma$ is particularly effective in reducing the $t\bar{t}$ background. Finally, a cut
on the pseudorapidity difference between the $W$ and $Z$ of $|\eta(Z)-\eta(W)|<1.2$ is effective
in reducing the $WZ$ background, although this remains the largest
background after all cuts as shown in figure~\ref{fig:TCsens}(a).

The expected signal sensitivity is computed using the sum of the reconstructed $\rho_{TC}$ mass spectra for the signal and backgrounds, taking into account the statistical fluctuations for a given integrated luminosity. 
It is assumed that the  probability density function is Gaussian for the signal and exponential for the background. 
The sensitivity estimator is given by
 $S_{\cal{L}}=\sqrt{2 {\rm ln} (\cal{L}_{S+B}/ \cal{L}_{B})}$, where $\cal{L}_{S+B}$, the signal plus background hypothesis, and $\cal{L}_{B}$, the null
 hypothesis. The sensitivity is computed for each signal point
 and the resulting contour plot in [$m(p_{TC}), m(\pi_{TC})$]
 space is shown in
 figure~\ref{fig:TCsens}(a). 5$\sigma$ sensitivities are obtained for integrated luminosities starting from $3$~fb$^{-1}$, before accounting for systematic uncertainties. Including the expected systematic uncertainties due to the detector, $5 \sigma$ discovery is possible starting from 4~fb$^{-1}$ of data.

\end{document}